\documentclass{aa}
\usepackage{graphics,epsfig}
\usepackage{graphicx}
\usepackage{array}
\usepackage{longtable}
\usepackage{supertabular}
\usepackage{amssymb}
\begin{document}
   \title{3C~129 : The GMRT observations}

%  \subtitle{I. The data}

   \author{Dharam Vir Lal\inst{1}, \and
          A. Pramesh Rao\inst{1}%\fnmsep\thanks{Just to show the usage of the elements in the author field}
          }

   \offprints{Dharam Vir Lal}

   \institute{National Centre for Radio Astrophysics,
              Pune University campus, Ganeshkhind, Pune - 411 007, India. \\
              \email{dharam@ncra.tifr.res.in} \\
              \email{pramesh@ncra.tifr.res.in} \\
             %\thanks{The university of heaven temporarily does not accept e-mails}
             }

   \date{Received ; accepted }

   \abstract{
We present radio maps of the head-tail radio galaxy, 3C~129
at 240 and 610 MHz using the Giant Metrewave Radio Telescope.
We have studied in detail the
morphology of the source and the distribution of spectral index over the object.
We find large-scale spectral steepening
along the jet away from the core.
Using synchrotron spectral ageing theory, we infer the
age (time since particle acceleration) of electrons at
various locations along the jet and ascribe an age of
$\sim$200~Myr to 3C~129.
We find that the ``Crosspiece" has a spectral index of
${\sim}-1$, similar to the jet, which is not consistent with the
suggestion by Lane et~al. (2002) that it could be a radio relic.
We have  also studied 3C~129.1 and the other sources in the field.
   }

 \authorrunning{Lal, D.V. \& Rao, A.P.}
 \titlerunning{GMRT observations of 3C 129}
   \maketitle

\section{Introduction}
Head-tail radio sources, which are almost always associated
with clusters of galaxies, are characterised by a highly
elongated radio structure with the associated optical
(usually giant elliptical) galaxy at one end.
A classical example of this class is 
3C~129 which, along with its companion 3C~129.1
(a wide-angle-tail),  is a member of the X-ray cluster 4U~0446+44
($z$ = 0.021) that happens to lie in the galactic plane
($b = 0^{\circ}.5, l = 160^{\circ}$) (\cite{Macdonaldetal1966} 1966,
\cite{HillLongair1971} 1971). 3C~129 is classified as a
prototype narrow-angle-tail (\cite{Laneetal2002} 2002) and has
a total flux density of
$\sim$6.27~Jy at 1400~MHz (\cite{Bolognaetal1969} 1969).
This very long narrow tailed source has been studied
at various wavelengths and with different
resolutions (\cite{Owenetal1979} 1979, \cite{Downes1984} 1984,
\cite{PerleyEricson1979} 1979, \cite{RudnickBurns1981} 1981 and
\cite{JagersdeGrijp1983} 1983). Recently this source was 
studied by \cite{Laneetal2002} (2002) at 330~MHz which is the lowest
radio frequency at which it has been observed.

A variety of models have been proposed to explain the morphology and
spectral index distribution of the head-tail sources and 3C~129 in particular
(\cite{JaffePerola1973} 1973, \cite{PacholczykScott1976} 1976
and \cite{Krawczynskietal2003} 2003).
In spite of possible systematic errors due to differing
UV coverages and possible artifacts introduced by deconvolution
procedures, meaningful conclusions have been  drawn about the
morphology of 3C~129 and its spectral index distribution.    
The main result is that
the radio spectrum has been found to steepen with distance
($S_\nu \propto \nu^{\alpha}$,
where $S_\nu$ and $\alpha$ are the flux density at frequency
$\nu$ and the spectral index respectively)
along the tail of 3C~129 (\cite{JaffePerola1973} 1973,
\cite{Miley1980} 1980 and \cite{vanBreugelandWillis1981} 1981)
and has been interpreted in terms of ageing of electron population.
A variant of the \cite {PacholczykScott1976} (1976) model
was suggested by \cite{PerleyEricson1979} (1979), who 
suggested that the longer wavelength
spectrum (between 74 and 1415~MHz) in any particular
location of the tail $\gtrsim$6$^\prime$ from the head,
is straight (a single power law) but
steadily steepens with increasing distance from the head.
Both these observations can well be explained using
a combination of continuous injection of new particles
(thus resupplying the high-energy electrons) and
particle diffusion out of the source (thus preventing
accumulation of low-energy particles). This led to the
refined Kardashev-Pacholczyk
(\cite{Kardashev1962} 1962, \cite{Pacholczyk1970} 1970)
and Jaffe-Perola (\cite{JaffePerola1973} 1973) models.

Recent high resolution observations by \cite{Laneetal2002} (2002)
at 330~MHz have shown a small perpendicular object, referred
to as the `Crosspiece', near the head of the galaxy.
This feature has a steep spectrum ($\alpha^{610}_{330}$ = $-$3;
$-1.5$ $<$ $\alpha^{1400}_{330}$ $<$ $-$1.1) and is
interpreted by them as a pre-existing fossil radio source that
was revived due to 3C~129 ploughing through it. The relic
source is compressed by the bow shock of 3C~129 causing
it to radiate behind the shock front, producing the
characteristic shape. If this interpretation is correct,
structures like the crosspiece should be rare which is verifiable
by making high resolution low frequency
maps of a well defined sample of head-tail sources.

We have started a project of studying a sample of
head-tail sources at 240 and 610~MHz using the GMRT.
In this paper we present results for 3C~129 and
describe the morphological and spectral properties.
We use the data to study  the behaviour of the
low frequency spectra along the tail and
compare it with the existing models. We confirm the existence
of some of the steep spectrum background sources reported by
\cite{Laneetal2002} (2002) at 330~MHz.

\section{Observations}
\label{obser}

\begin{table}
\caption{Observing log for 3C~129. Centre of the field
was at RA$_{\rm B1950}$ = 04 45 25.06 and Dec$_{\rm B1950}$ = 44 57 00.09
and use 86.5 Mpc (H$_0$ = 71 km~s$^{-1}$~Mpc$^{-1}$) as the
distance to 3C~129.}
\centering
\begin{tabular}{l|cc}
\hline \\
 & 610 MHz & 240 MHz \\
\\
\hline \\
Observing date & 24 May~2002 & 19 May~2002 \\
Duration & 7.45 Hrs & 6.84 Hrs \\
Centre frequency & 613.375~MHz & 242.062~MHz \\
Nominal bandwidth & 16~MHz & 8~MHz \\
Effective bandwidth & 12.75~MHz & 5~MHz \\
Primary beam & 43$^{\prime}$ & 108$^{\prime}$ \\
Synthesized beam & 8.0$^{\prime\prime}$ $\times$ 7.1$^{\prime\prime}$  &
               13.1$^{\prime\prime}$ $\times$ 10.8$^{\prime\prime}$ \\
~~~~~~~~~~~~~~~(P.A.) & 36$^{\circ}$.4 & 35$^{\circ}$.2 \\
Sensitivity ($\sigma$)& 0.3~mJy~beam$^{-1}$ & 1.3~mJy~beam$^{-1}$ \\
Dynamic range & $\sim$440 & $\sim$410 \\
Calibrator & 3C~147 & 3C~147 \\
$S_{\nu}$ (3C~147) & 38.26 Jy & 59.14 Jy \\
\\
\hline
\end{tabular}
\label{log}
\end{table}
We made full synthesis observations of 3C~129 at 240 and 610 MHz
using the GMRT on 19 and 24 May~2002 respectively in the standard
spectral line mode with a spectral resolution of 128 kHz.
Table~\ref{log} gives the details of the observations.
The visibility data were converted to FITS and analysed using standard AIPS.  
The flux calibrator 3C~147 was also the amplitude and phase calibrator
and was observed once every 40~min.
3C~147 was used to estimate and correct for the bandpass shape and 
to set the flux density scale which is an extension of the
\cite{Baarsetal1977} (1977) scale to low frequencies,
using the coefficients in the AIPS task `SETJY'.
The error in the estimated flux density,
both due to calibration and systematic, is $\lesssim$ 5\%.
In addition to normal editing of the data, the
observations at 240 MHz were affected by intermittent
radio frequency interference (RFI) and
channels affected due to RFI were identified and edited, 
after which the central channels were averaged using `SPLAT' to reduce the
data volume. To avoid bandwidth smearing, 5~MHz of clean band at 240~MHz 
was reduced to 4 channels of 1.25~MHz each.
At 610~MHz where there was little RFI, 12.75~MHz of
clean band was averaged to give 2 channels of 6.375~MHz each.

While imaging, 37 facets (obtained using AIPS task `SETFC'),
spread across $\sim$2$^\circ\times2^\circ$ field were used at 240~MHz and
5 facets covering  slightly less
than 0.$^\circ7\times0.^\circ7$ field, were used at 610~MHz to map
the field using the AIPS task `IMAGR'. 
We used `uniform' weighting and the 3$-$D option for W~term
correction throughout our analysis.
The presence of a large number of point sources in the field
allowed us to do phase self-calibration to improve the image.
After 2-3 rounds of phase self-calibration, a final self-calibration 
of both amplitude and phase was made to get the final image.
At each round of self-calibration, the image and the visibilities
were compared to check for the improvement in the source model.
The final maps were combined using `FLATN' and corrected for
the primary beam of the GMRT antennas.

\section{Radio maps}
\label{rad_map}

\begin{figure*}[ht]
\begin{center}
\begin{tabular}{l}
\includegraphics[width=17.5cm,angle=-180]{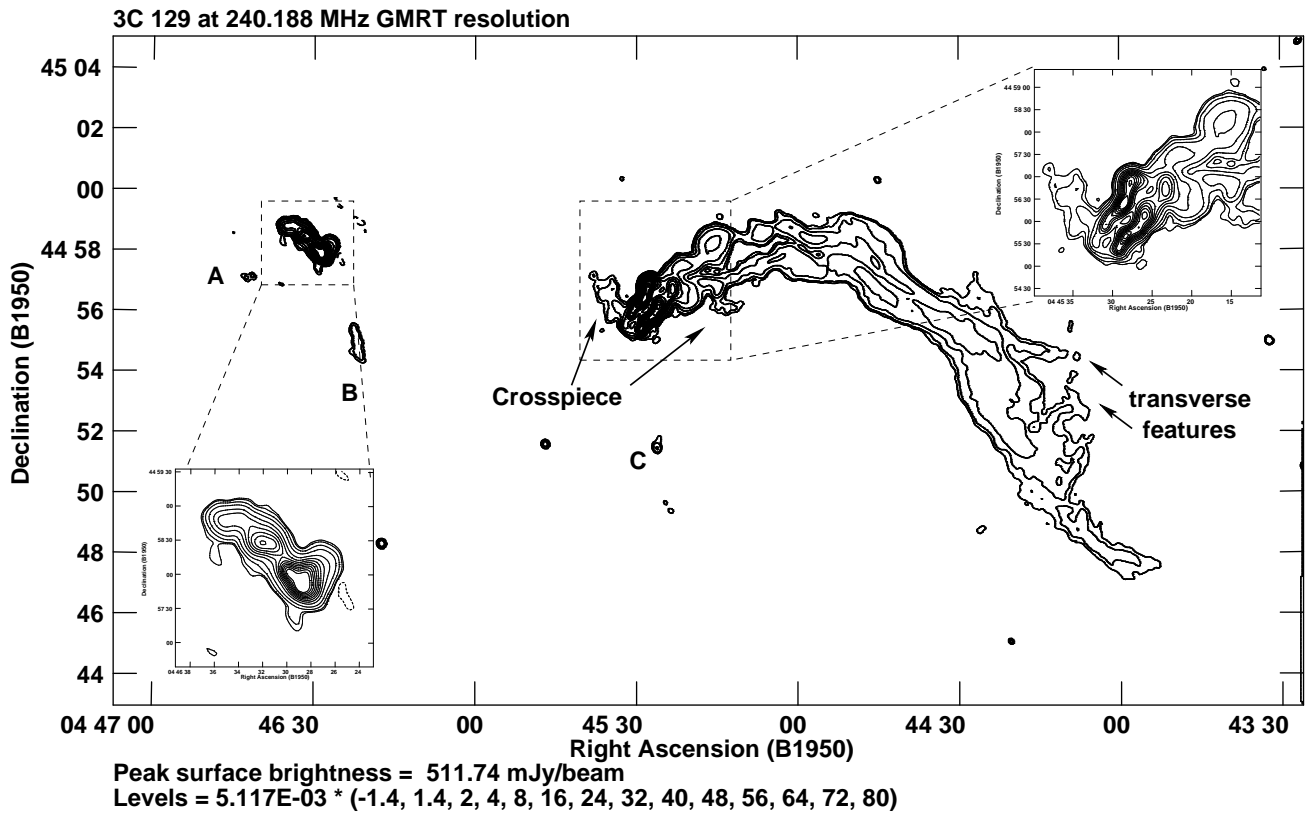}
\end{tabular}
\end{center}
\caption{Full synthesis GMRT map of 3C 129 and the sources
in its field at 240 MHz.
The two insets show 3C~129.1 (lower left) and head of 3C~129 (upper right).
The CLEAN beam is 13$^{\prime\prime}$.1~$\times$~10$^{\prime\prime}$.8
at a P.A. of 35$^{\circ}$.2.}
\label{240mhzmap}
\end{figure*}

\begin{figure*}[ht]
\begin{center}
\begin{tabular}{l}
\includegraphics[width=17.5cm,angle=-180]{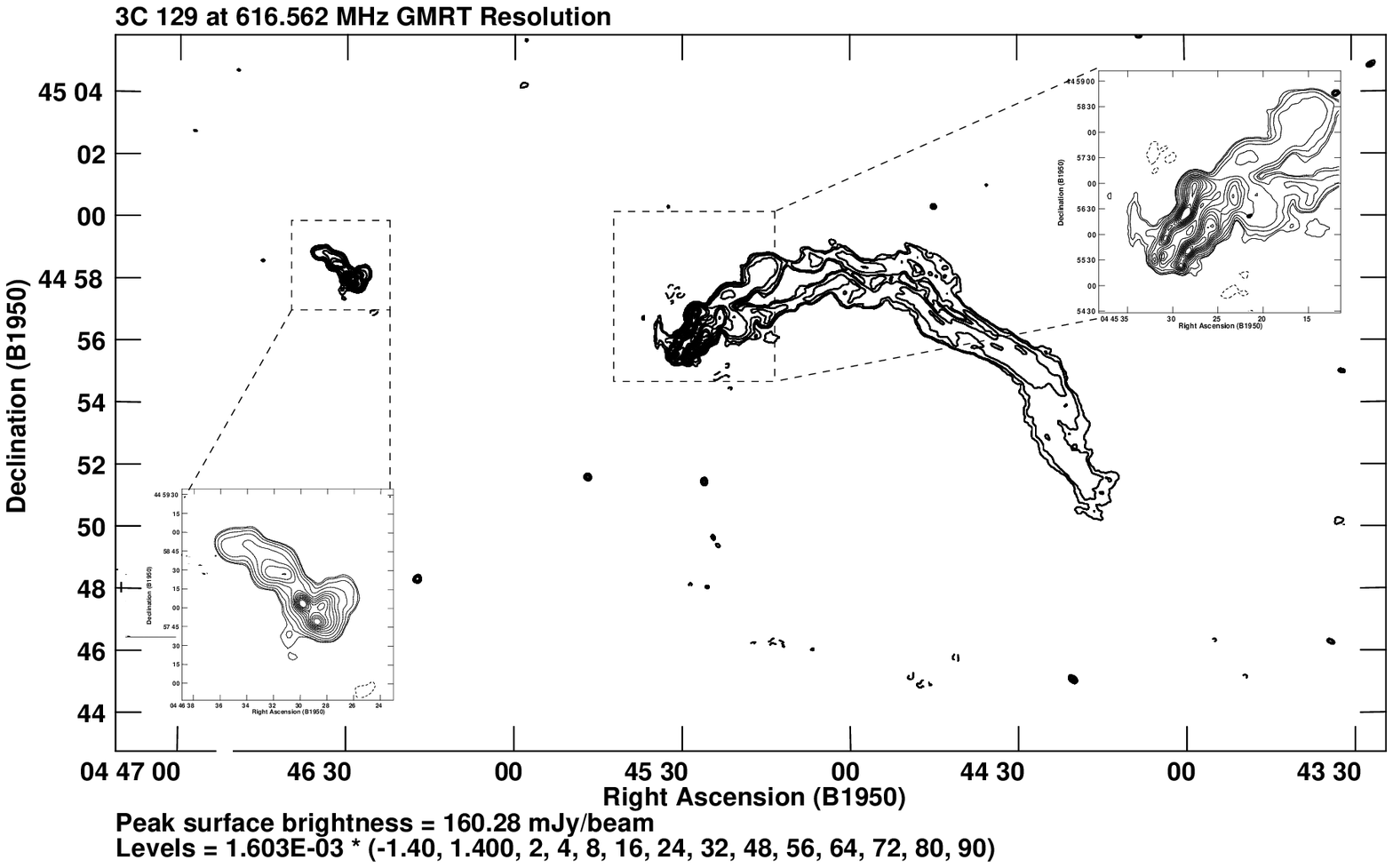}
\end{tabular}
\end{center}
\caption{Full synthesis GMRT map of 3C 129 and the sources
in its field at 610 MHz.
The two insets show 3C~129.1 (lower left) and head of 3C~129 (upper right).
The CLEAN beam is 8$^{\prime\prime}$.0~$\times$~7$^{\prime\prime}$.1
at a P.A. of 36$^{\circ}$.4. The `cross' in the lower right
inset marks the position of the optical nucleus with the size
of the cross representing 1$\sigma$ uncertainty in position.}
\label{610mhzmap}
\end{figure*}

The high angular resolution map of 3C~129
at the lowest frequency of 240~MHz is shown in
Fig.~\ref{240mhzmap}. Fig.~\ref{610mhzmap} shows
the image at 610~MHz. The images shown
have nearly complete UV coverage, an angular resolution
12$^{\prime\prime}$ and 7$^{\prime\prime}$ and the
rms~noise in the maps are $\sim$1.3~mJy~beam$^{-1}$ and
$\sim$0.3~mJy~beam$^{-1}$ at 240 and 610~MHz respectively.
The dynamic range in each of the maps is $\gtrsim$400 (see also
Table~\ref{log}).
The GMRT has a hybrid configuration (\cite{Swarupetal1991} 1991
and \cite{AnanthRao2002} 2002)
with 14 of its 30 antennas located in a central compact array
with size $\sim$1.1~km and the remaining antennas distributed in
a roughly `Y' shaped configuration, giving a maximum baseline length
of $\sim$25~km.
The hybrid configuration gives reasonably good sensitivity
for both compact and extended sources.
The baselines obtained from antennas in the central
square are similar in length to those of the VLA~$D$-array,
while the baselines between the arm antennas are
comparable in length to the VLA~$B$-array. A single
observation with the GMRT hence yields information on both
small and large angular scales.
To enhance the low surface brightness features,
the maps were therefore made at various scales.
The morphology and spectra of 3C~129 and some of the
other sources in the field are described below.
In the discussion we have also made use of the VLA 330~MHz
image of the field kindly supplied by W.M. Lane and published
in \cite{Laneetal2002} (2002). Since, maps published in
\cite{Laneetal2002} (2002) are in B1950 epoch, we also
present all our maps in the same epoch.

\subsection{Radio morphology of 3C~129}
\label{rad_map3c129}

This complex radio source shows a twin-tailed morphology
at both frequencies.
The maps illustrate the main characteristics of
3C~129 at low frequencies.
The curving tail is at least 22$^{\prime}$ long at 240~MHz
(Fig.~\ref{240mhzmap}), while at the higher frequency
(610 MHz,  Fig.~\ref{610mhzmap}) the tail is detectable out to
$\sim$20$^{\prime}$ from the head. This is because the spectrum steepens
along the length of the tail and the extended outer
components are not bright enough to be detected
in high frequency observations (\cite{vanBreugel1982} 1982).
The nuclear source of 3C~129 is unresolved at both frequencies.
In both the radio maps, dissimilar behaviour of the two
jets indicates that the upper jet is brighter than the lower.
Using the radio morphology of them at 240
and 610 MHz, we further find that in the upper jet,
the wiggles continue and lead to the far tail, while
the lower jet ends earlier and disappears at a distance
of $\sim$7$^{\prime}$ from the nucleus.
There is also a sharp drop in surface brightness at this
position.

In Table 2, we show the integrated flux densities of 3C~129 and other
extended sources in the field along with previous measurements at
frequencies below 1500~MHz. Our estimate at 610~MHz agrees well with
that of \cite{Jagers1987} (1987) got using the WSRT while our 240~MHz
is in reasonable agreement with what is expected from its spectrum.
We therefore believe that we have not
lost any flux density in our interferometric observations.

\begin{table*}
\caption{The total intensity comparisons for 3C~129 and field sources.
The flux densities quoted are in Jy along with corresponding errorbars
(1$\sigma$). The 178~MHz measurements are using
large Cambridge interferometer (\cite{Ryle1960} 1960), 325 \& 1400~MHz
measurements are using VLA (\cite{Laneetal2002} 2002 \&
\cite{Condonetal1998} 1998), and 240 \& 610~MHz measurements are
our observations; and
$^{\P}$: \cite{PerleyEricson1979} 1979,
$^{\S}$: \cite{Bennett1962} 1962 (the revised 3rd Cambridge catalogue,
the flux integrated from map for 3C~129 and 3C~129.1 is 46.9$\pm$4.7~Jy
(\cite{Goweretal1967} 1967)),
$^{\dag}$: \cite{Jagers1987} 1987, and
$^{\ddag}$: \cite{WhiteBecker1992} 1992 (Green Bank, Northern Sky Survey).
We also quote 3$\sigma$ values for the non-detections ({\it e.g.}
sources, A \& B at 610~MHz, and Crosspiece, A, B \& C at 1400~MHz).}
\centering
\begin{tabular}{r|ccrrrl}
\hline \\
   &74 MHz& 178~MHz & 240~MHz & 330~MHz & 610~MHz & 1400~MHz \\
   & \multicolumn{6}{c}{Flux density (Jy)} \\
\\
\hline \\
3C~129 &90.0$\pm$5.0$^{\P}$ &21.0$^{\S}$&27.5$\pm$1.3 &21.2$\pm$1.4~& 14.5$\pm$1.4~&7.0$\pm$1.0 \\
        & & & & &14.4$\pm$0.2$^{\dag}$ & 5.3$^{\ddag}$ \\
3C~129.1&&11.0$^{\S}$& 4.6$\pm$0.4~& 3.5$\pm$0.8~&2.0$\pm$0.5~ & 0.9$\pm$0.4 \\
Crosspiece & & &0.032$\pm$0.005 & 0.020$\pm$0.003&0.012$\pm$0.002 & $<$~0.003 \\
     A & & &0.05$\pm$0.002 & 0.03$\pm$0.002 & $<$~0.003 & $<$~0.003 \\
     B & & &0.14$\pm$0.005 & 0.05$\pm$0.001 & $<$~0.003 & $<$~0.003 \\
  C & & & 0.040$\pm$0.003 & 0.03$\pm$0.004 & 0.014$\pm$0.003 & $<$~0.003 \\
\\
\hline
\end{tabular}
\label{flux_density}
\end{table*}

\subsection{Radio morphology and integrated spectra
of sources in the 3C~129 field}
\label{other}

The 20$^{\prime}$ region around 3C~129 has some
interesting sources, like 3C~129.1, the crosspiece,
etc., which we describe here.

\paragraph{\underline {3C~129.1}}

Our observations of 3C~129.1 show it to be an elongated radio
source with angular size of $\sim$2$^{\prime}$ along the north-east.
High resolution observations at 2.7~GHz by \cite{Downes1980} (1980)
suggest it to be a double source with a prominent core. The core and
double structure are also seen in our 610~MHz high resolution map. But
at both 240 and 610~MHz, there is considerable diffuse emission around the
source, some of which extends well beyond the double structures, typical of
an Fanaroff-Riley type I (FR~I) source (\cite{FanaroffRiley} 1974).
The diffuse emission shows some curvature which is consistent
with the suggestion by \cite{Laneetal2002} (2002) that it could be a
wide-angle-tail source with the source moving with small transverse
velocity, though it could also be due to buoyancy effects
(see Sect.~\ref{discuss}).

This radio source has a total flux density of $\sim$0.94~Jy at
1400~MHz (\cite{Condonetal1998} 1998) and our maps at 240 and 610~MHz
give a total flux density of 4.6 and 2.0~Jy respectively.
Its low frequency spectral index is,
$\alpha_{240}^{610}$ = $-$0.86$\pm$0.25 which is typical of
sources associated with clusters (\cite{BaldwinScott1973} 1973).
There is a weak gradient in spectral index behaviour along the
source (Fig.~\ref{sp_in}); with centre which being flat is
surrounded by steeper spectrum region.

\paragraph{\underline {The Crosspiece}}

Using WSRT, \cite{JagersdeGrijp1983} (1983) at 608~MHz
reported a small projection near the head on one side of 3C~129.
This projection has been confirmed by \cite{Laneetal2002} (2002)
in their high resolution map at 330~MHz who have named it as the
`Crosspiece'.
We also detect this feature clearly at both the observed frequencies
and see evidence for this feature in the NVSS at 1400~MHz.
From its integrated flux densities, we estimate its spectral index
$\alpha_{240}^{610}$ as $-$1.04$\pm$.22 and is roughly straight
between 240 and 1400~MHz. This value is flatter than the
spectral index ($\alpha_{330}^{600}$ of $-$3.0) estimated
by \cite{Laneetal2002} (2002), who had used the
results of \cite{JagersdeGrijp1983} (1983) at 608~MHz and
they seem to have underestimated its flux density.

\paragraph{\underline {Sources A, B and C}}
The faint sources A and B are seen in VLA maps at
74~MHz and 330~MHz (\cite{Laneetal2002} 2002). We
detect sources A and B at 240~MHz, but not at 610~MHz.
Source C has been seen in NVSS (\cite{Condonetal1998} 1998) and
at 330~MHz (\cite{Laneetal2002} 2002). We detect source C
at both 240 and 610 MHz. It has an integrated flux density
of 40.4 and 13.9~mJy at 240 and 610~MHz respectively.
These detections of source~C, give spectral indices
$\alpha_{240}^{610}$ and $\alpha_{240}^{1400}$ as
$-$1.14$\pm$0.31 and $-$1.51$\pm$0.07 respectively.
Using detection at 240 MHz and upper detection limits
at 610~MHz for the sources A and B, we conclude that
they are steeper than, $\alpha_{240}^{610}$,
$-$3.0$\pm$0.3 and $-$4.1$\pm$0.3 respectively.

\section{Low frequency spectra of 3C~129}
\label{sour_spec}

The observations and morphology described in the previous section
(Sect.~\ref{rad_map3c129}) allow us to investigate in detail the spectral
index distribution of 3C~129.
Both the GMRT 610~MHz and VLA 330~MHz
maps were restored with the restoring beam corresponding to
the 240~MHz map (13$^{\prime\prime}$.1~$\times$~10$^{\prime\prime}$.8 at P.A.
35$^{\circ}$.2). The final calibrated UV data at 610~MHz
was first mapped using UV~taper 0--22~k$\lambda$,
which is similar to that of
240~MHz data and then restored using the restoring beam
corresponding to the 240~MHz map. At 330~MHz,
since we had only map, we restored it using
restoring beam corresponding to that of 240~MHz map.
These three maps restored with the same resolution,
GMRT 240 \& 610~MHz and VLA 330~MHz,
were used for the spectral analysis.
We determine the spectral index distribution using a
number of techniques.

\begin{figure}
\begin{center}
\begin{tabular}{l}
\includegraphics[width=6cm,angle=270]{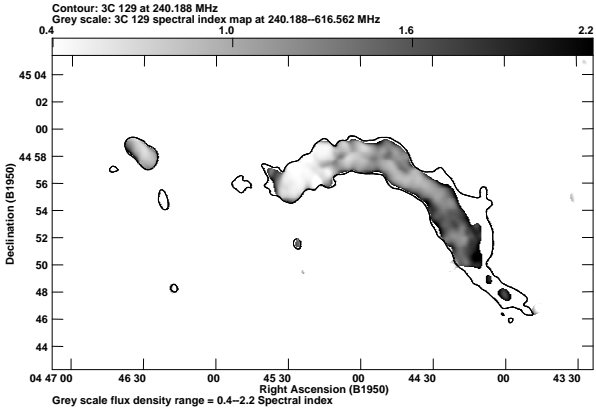}
\end{tabular}
\end{center}
\caption{Distribution of the spectral index $\alpha$
($S_{\nu} \propto \nu^{\alpha}$)
between 240 \& 330 MHz (upper) and 240 \& 610 MHz (lower)
as computed from images made with a
matched restoring beam (36$^{\prime\prime}$.9~$\times$~28$^{\prime\prime}$.4
at P.A. = 21$^{\circ}$.6). The lighter regions represent the
relatively flat spectrum regions as compared to the darker
regions which represent steep spectrum (although the full range
of spectral index is $-$3.3 to 2.4, we have shown only 0.4 to 2.2
for clarity).
Also, plotted is the 3$\sigma$ contour corresponding to map at
240~MHz. The blank region inbetween closed contour and grey scale
is probably steeper (or intermediate) than the shown darker
regions in the grey scale plot.}
\label{sp_in}
\end{figure}

The standard, direct method of determining the spectral
index between maps $S_{\nu_1}(x,y)$ and $S_{\nu_2}(x,y)$
at two frequencies ${\nu_1}$ and ${\nu_2}$ is given by
$$
\alpha_{\nu_1,\nu_2}(x,y) \equiv
\frac{{\rm log}~(S_{\nu_1}/S_{\nu_2})}{{\rm log}~(\nu_1/\nu_1)}.
$$
We also looked at the 
spatially `filtered' images (\cite{Wrightetal1999} 1999)
and `spectral tomography' images (\cite{Rudnick2002} 2002)
to enhance weak features seen in the maps.
We do not find evidence for a new feature emerging in
the filtered or tomographic images, which could have been
missed in the single frequency maps.
While, it could also be due to sensitivity limits of our maps,
this is possibly due to the fact that these techniques
are effective in separating overlapping large-scale and
small-scale features with different spectral distributions.
Hence, these techniques are found to be useful for radio
galaxies, supernova remnants, etc. (\cite{Rudnick2002} 2002).
Whereas, head-tail sources are understood to be
FR~I radio sources moving through the
gas in the cluster and the shape of the source is due to the
diffuse radio emitting plasma being decelerated by the
intracluster medium (ICM) (\cite{Mileyetal1972} 1972).
Such sources do
not suffer from regions with overlapping features with
different spectral indices and therefore these techniques
did not seem effective.
However, in these series of slices through the source (in
other words, spectral tomography images for a range of spectral
indices), different parts of the source become prominent
in different sets of images and provide additional
support to their being real features.

\begin{figure}
\begin{center}
\begin{tabular}{l}
\includegraphics[width=6cm,angle=270]{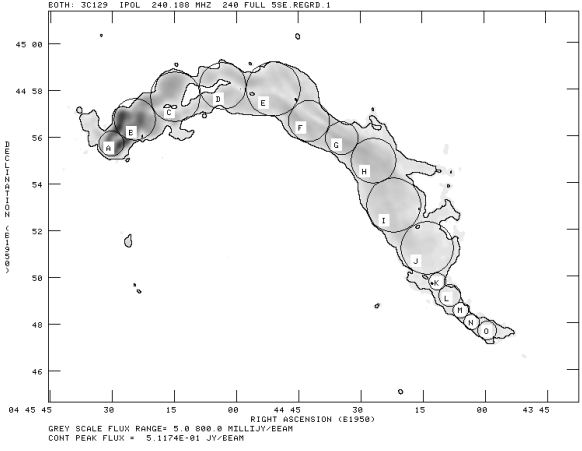}
\end{tabular}
\end{center}
\caption{Figure showing sampled 15 distinct regions
along the ridge line of maximum emission. The size of
each region correspond to the width of the source at
the position of the region. These regions (marked, A, B, C, ... and O)
are different from the field sources A, B and C discussed in
previous section.}
\label{rep_fea}
\end{figure}

\begin{figure}
\begin{center}
\begin{tabular}{l}
\includegraphics[width=8.7cm]{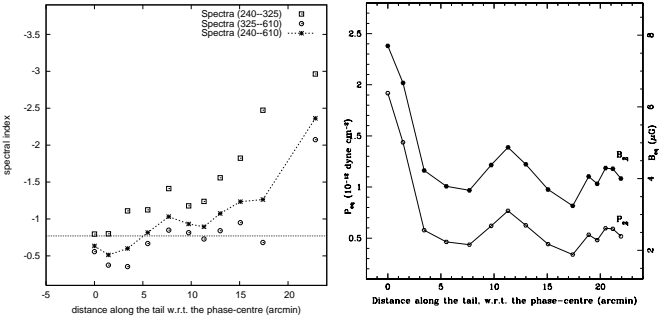}
\end{tabular}
\end{center}
\caption{Left: Spectral index profile along the 3C~129.
The profile corresponds to the regions shown in Fig.~\ref{rep_fea}.
The horizontal line shows the averaged spectral index
($-$0.77) using integrated flux density of the source.
A single averaged value of spectral index positioned at the
position of O is shown for all 5 regions,
from K to O, which are
at the farther end of the tail. Errorbars (1$\sigma$) on the spectral
indices are between 0.03 and 0.22.
Right: The minimum pressure (dyne~cm$^{-2}$)
and equipartition magnetic field ($\mu$G)
of the radio plasma along the tail at the sampled positions.}
\label{profiles_1}
\end{figure}

In order to set more reliable estimates of the large scale behaviour
of the source, we selected 15 circular regions (A, B, C, ... and O),
centered on the ridge lines of maximum emission
of diameters corresponding
to the width of the source at that position.
The position and size of these regions are shown in
Fig.~\ref{rep_fea}.
For each of these regions, we compute the total
flux and construct the two frequency radio
energy spectrum as a function of distance with respect
to the head.
Table~\ref{cross_head_eq} gives brightness (in units of Jy~arcmin$^{-2}$)
at 240, 330 and 610~MHz for each of these regions.
The distances in Col~2 correspond to the distance of each of the
regions, in arcmin, from the head
along the tail. The profiles of the two frequency spectral
indices, as a function of the distance from the
head are shown in Fig.~\ref{profiles_1}.

Figs.~\ref{sp_in} (spectral index distribution) and
~\ref{profiles_1} (spectral index profile) shows that
the spectra is flattest ($\alpha$  $\gtrsim$ $-$0.5)
at the core of 3C~129 in both, $\alpha_{240}^{610}$
\& $\alpha_{240}^{330}$, spectral index maps.
Also in the spectral index map
at 240 and 610~MHz (Fig.~\ref{sp_in}), the
spectral index flattens slightly along the jet to a minimum
which occurs around 2$^{\prime}$ (see Fig.~\ref{profiles_1}).
Between 2$^{\prime}$ and 7$^{\prime}$ from the head,
the spectrum of the source steepens
from $-$0.5 to $-$0.8; from 7$^{\prime}$ to 12$^{\prime}$
the spectrum remains constant; and
from 12$^{\prime}$ up to the farthest tail end, spectral index
steepens further from $-$0.8 to $\sim$$-$2.0.
Whereas, in the low-frequency spectral index profile
at 240 and 330~MHz, the region up to 2$^{\prime}$ from core
has a constant spectral index of $-$0.8;
between 2$^{\prime}$ and 7$^{\prime}$ from the head,
the spectrum of the source steepens from $-$0.8 to $-$1.2;
from 7$^{\prime}$ to 12$^{\prime}$
the spectrum remains constant;
and from 12$^{\prime}$ to the farthest tail end, spectral index
steepens further from $-$1.2 to $\sim$2.5.
Thus, the spectrum from our observations, roughly supports
\cite{PacholczykScott1976} (1976) interpretation.

In terms of the electron spectrum between 240 and 610~MHz,
the faint steepening of electron spectral index, $\gamma$
(N(E)~dE~$\propto$~E$^{-\gamma}$~dE and
$\gamma$~=~1~$-$~2$\alpha$)
is observed from the head towards the tail.
The electron spectrum steepens from 2 to 2.6 from region
close to the core and up to $\sim$7$^{\prime}$; from
7$^{\prime}$ to 12$^{\prime}$ the electron spectrum remains
constant; and finally beyond 12$^{\prime}$ results in even
more steepening of the electron spectrum. This steepening
of electron energy spectrum is due to synchrotron ageing
of the radio plasma. The lowest energy electrons
diffuse into the lower pressure regions and form the
steep spectrum tail at the farther end from the head.
This may be quite easy along the well ordered magnetic
field of 3C~129 as suggested by \cite{vanBreugel1982} (1982).

Furthermore, the spectra of each of the distinct regions 
(Fig.~\ref{rep_fea})
using the data (Table~\ref{cross_head_eq}) is straight.
Even on adding the upper detection limits at 1400~MHz
using published map by \cite{vanBreugel1982} (1982), the spectra
of each of these components remain roughly straight.
In other words, our data provides support to the notable
result brought out by \cite{PerleyEricson1979} (1979) that
straight spectra are seen at all points down the tail.

\section{Physical parameters of the source}
\label{physical}
Using the standard formulae (\cite{Pacholczyk1970} 1970),
we estimate the equipartition internal
energy density (U$_{\rm min}$), pressure
(P$_{\rm eq}$ = U$_{\rm min}$/3), and magnetic energy
(B$_{\rm eq}$)
for the sampled regions in 3C~129
(see Table~\ref{cross_head_eq}).
We assume that the transverse size of each component in
the source is the path length through the source along
the line of sight.

Fig.~\ref{profiles_1} shows the minimum
pressure and equipartition magnetic field of 3C~129
plotted as a function of distance along the tail
from the head. It therefore seems that the pressure
in farther end of the tail is
less by a factor of 4 as compared to the region
close to core and hence is more diffuse.

\subsection{Source age}
\label{spec_mod}

The observed shape of the radio spectrum results from competition
between energy injection and losses due to adiabatic
expansion, synchrotron emission and inverse Compton (IC)
scattering off CMB photons. Therefore, the shape of the
radio spectrum along with a detailed
understanding of these processes can yield the age of the
source, the duration of initial particle injection and the
magnetic field strength.

\begin{table*}[ht]
\caption{Distance from the head along the tail (in arcmin),
size (in arcmin), flux density of all the distinct
regions at 240, 330 and 610 MHz (in Jy~arcmin$^{-2}$), and
equipartition parameters at the sampled positions.
The age (expressed in Myr) is calculated following formalism in
Sect.~\ref{spec_mod} and using $\nu_{\rm br}$ of 610~MHz
(if $\nu_{\rm br}$ = 1400~MHz, these would drop by a factor of
$\sim$1.5).}
\centering
\begin{tabular}{l|crrrrcccr}
\hline \\
 & distance
 & size & $B_{240~{\rm MHz}}$ & $B_{330~{\rm MHz}}$ & $B_{610~{\rm MHz}}$ 
 & U$_{\rm min}$ & P$_{\rm eq}$ & B$_{\rm eq}$ & Age \\
 & ($\prime$) & ($\prime$)
      & \multicolumn{3}{c}{(Jy~arcmin$^{-2}$)} 
 & (erg~cm$^{-3}$) & (dyne~cm$^{-2}$) & ($\mu$G) & (Myr) \\
\\
\hline \\
A& 0.00 & 1.07 & 4.444 & 3.492 & 2.458 
&5.6 $\times$ 10$^{-12}$ &1.9 $\times$ 10$^{-12}$ & 7.7 & ~58.7 \\
B& 1.44 & 1.80 & 4.527 & 3.553 & 2.806 
&4.3 $\times$ 10$^{-12}$ &1.4 $\times$ 10$^{-12}$ & 6.7 & ~71.3 \\
C& 3.42 & 2.13 & 1.089 & 0.778 & 0.622 
&1.7 $\times$ 10$^{-12}$ &5.8 $\times$ 10$^{-13}$ & 4.2 & 125.3 \\
D& 5.51 & 2.00 & 0.694 & 0.494 & 0.324 
&1.4 $\times$ 10$^{-12}$ &4.6 $\times$ 10$^{-13}$ & 3.8 & 141.2 \\
E& 7.68 & 2.33 & 0.728 & 0.474 & 0.278 
&1.3 $\times$ 10$^{-12}$ &4.4 $\times$ 10$^{-13}$ & 3.7 & 145.7 \\
F& 9.73 & 1.73 & 1.000 & 0.700 & 0.420 
&1.9 $\times$ 10$^{-12}$ &6.2 $\times$ 10$^{-13}$ & 4.4 & 120.4 \\
G&11.31 & 1.40 & 1.174 & 0.807 & 0.510 
&2.3 $\times$ 10$^{-12}$ &7.7 $\times$ 10$^{-13}$ & 4.9 & 106.2 \\
H&12.99 & 1.93 & 1.134 & 0.707 & 0.417 
&1.9 $\times$ 10$^{-12}$ &6.3 $\times$ 10$^{-13}$ & 4.4 & 119.7 \\
I&15.08 & 2.33 & 0.744 & 0.428 & 0.236 
&1.3 $\times$ 10$^{-12}$ &4.4 $\times$ 10$^{-13}$ & 3.7 & 144.8 \\
J&17.40 & 2.27 & 0.457 & 0.216 & 0.141 
&1.0 $\times$ 10$^{-12}$ &3.4 $\times$ 10$^{-13}$ & 3.2 & 164.7 \\
K&18.90 & 0.73 & 0.325 & 0.146 & 0.012 
&1.6 $\times$ 10$^{-12}$ &5.3 $\times$ 10$^{-13}$ & 4.1 & 131.0 \\
L&19.70 & 0.93 & 0.345 & 0.132 & 0.027 
&1.4 $\times$ 10$^{-12}$ &4.8 $\times$ 10$^{-13}$ & 3.9 & 138.5 \\
M&20.49 & 0.67 & 0.360 & 0.141 & 0.027 
&1.8 $\times$ 10$^{-12}$ &6.0 $\times$ 10$^{-13}$ & 4.3 & 123.1 \\
N&21.17 & 0.67 & 0.354 & 0.153 & 0.048 
&1.8 $\times$ 10$^{-12}$ &5.9 $\times$ 10$^{-13}$ & 4.3 & 123.7 \\
O&21.94 & 0.80 & 0.337 & 0.135 & 0.075 
&1.6 $\times$ 10$^{-12}$ &5.2 $\times$ 10$^{-13}$ & 4.0 & 133.1 \\
\\
\hline
\end{tabular}
\label{cross_head_eq}
\end{table*}

Following \cite{Miley1980} (1980),
we derive the upper limit to the age of the synchrotron
electrons at several positions along the radio tail.
The total radiative age is calculated~as
$$
{\rm t} = 1060 \frac{{\rm B}^{0.5}}{({\rm B}^2 +
{\rm B}_{\rm IC}^2)}[(1 + z)\nu_{\rm br}]^{-0.5}~{\rm Myr},
$$
where $\nu_{\rm br}$ is expressed in GHz,
and is the break frequency of the injected
electron population, where the spectrum changes by 0.5.
The magnetic fields, B and B$_{\rm IC}$, are expressed
in $\mu$G and the radiative age, t, in Myr. We further
choose B~=~B$_{\rm IC}$/$\sqrt{3}$, the value that gives
maximum radiative lifetime (\cite{vanderLaanPerola1969} 1969)
and B$_{\rm IC}$ = 4~(1 + $z)^2$~$\mu$G.
With our sparse measurements for each of the regions,
the break frequency, if it exists at all in 3C~129,
must lie either above 610~MHz or below 240~MHz.
This is consistent with the larger extent suggested by
\cite{PerleyEricson1979} (1979), who suggested that the break
frequency must be either above 1400~MHz (and probably
above 2695~MHz, \cite{Riley1973} 1973) or below 43~MHz.
We give age derived using equipartition magnetic
field and using 610~MHz as the break frequency at
different locations along the source in Table~\ref{cross_head_eq}.
We find that age of each of the regions between `C' and `O'
is within $\sim$50~Myrs. The core is the youngest with
an age of $\sim$59~Myr.
The assumption that there is equipartition of energy
between the radiating particles and magnetic field seems
to break at regions close to `J', where the age peaks as
compared to the adjacent regions. Since, we have not
determined B$_{\rm eq}$ and $\nu_{\rm br}$ independently
(\cite{KomissarovGubanov1994} 1994 and \cite{Sleeetal2001} 2001) and
because the age of any region from 3$^{\prime}$ (region `C') onward
is not very different from the age of the farthest region,
labeled `O' in Table~\ref{cross_head_eq},
we use the averaged age, {\it i.e.} $\sim$124~Myr, of all the
regions from `C' to `O' as the
representative age of the 3C~129.
Assuming 1400~MHz (\cite{PerleyEricson1979} 1979)
as the break frequency, the age of 3C~129 is $\sim$86~Myr.
\cite{Mileyetal1972} (1972)
suggests the age of 3C~129 to be of the order of
$\sim$200~Myr, which is based on the length of the
tail and galaxy velocity. Therefore, processes like
particle acceleration or injection of new particles
into the tail must be present. On the other hand,
if the break frequency is $\sim$240~MHz, then the age
estimate of $\sim$200~Myr for 3C~129 is in
agreement with the age estimate of \cite{Mileyetal1972} (1972).
We understand that without knowing the history of the
relativistic electron population, {\it i.e.} their
current and past magnetic field environments, it is
difficult to assign a well defined age to 3C~129 on the
basis of spectral behaviour.
Nevertheless, we assume that equipartition magnetic field
holds good at most of regions along the jet.
The age of $\sim$200~Myr therefore seems plausible for 3C~129
as suggested by \cite{Mileyetal1972} (1972) and
break frequency occurs at $\sim$240~MHz.

\section{Discussion}
\label{discuss}

The low frequency observations combined with recent observations at other 
wavelengths raise a number of questions regarding
the nature and origins of the various sources in the field.
Some of the earlier scenarios that were based on limited information 
need to be revised. In this section we will discuss the implications 
of the new observations on our understanding of the source.

\paragraph{\underline {Nature of the Crosspiece:}}
The integrated radio flux densities of relic sources fall
very rapidly with frequency, with power-law slopes, $\alpha$,
between $-$2.1 and $-$4.4 near 1400~MHz (\cite{Sleeetal1994} 1994).
Their break frequency is below 100~MHz with typical age of
$\sim$300~Myr ({\it e.g.} archetype radio relic,
Abell~85; \cite{Sleeetal2001} 2001).
Although the crosspiece has low surface brightness,
we find that the spectral index of it is not very different
from usual radio plasma outflow of 3C~129 (Sect.~\ref{other}).
We also find that the spectral index, between 240 and 1400~MHz,
of it is roughly straight.
The scenario proposed by \cite{Laneetal2002} (2002)
in which the crosspiece is a steep spectrum relic source that has
been reactivated by 3C~129 plowing and compressing it 
seems to be untenable since the adiabatic compression process
cannot change the spectrum of the source. Close examination
of the tail of 3C~129 reveals features extending perpendicular
to the tail, similar to crosspiece, further out along the
tail (see transverse features close to regions `H' and `J', 
shown in Fig.~\ref{240mhzmap}). Such features seem to be an integral
part of the tail of 3C~129 and could arise due to perturbations 
in the jet and/or may be due to buoyancy effects because of
density inhomogeneities in ICM.
Furthermore, since the spectral index along the tail and at the
crosspiece are similar and if we assume, following
\cite{Laneetal2002} (2002), that the break frequency is at 1400~MHz,
the maximum radiative age of the crosspiece is $\sim$60~Myr.
This does not seems to agree with suggested dynamical age of
about 40~Myr (\cite{JagersdeGrijp1983} 1983).
Also, an age of 40~Myr along with an infall velocity of
1630~km~s$^{-1}$ (\cite{Laneetal2002} 2002)
for 3C~129, would put crosspiece at a distance of
at least 8$^{\prime}$.3 from the head.
Hence, crosspiece being one-sided ejection of plasmoids or beams,
which took place about 40~Myr (\cite{JagersdeGrijp1983} 1983)
is also unlikely.
We therefore believe that the features like crosspiece are a
manifestation of inhomogeneities in the medium and/or
perturbations in the jet.

\paragraph{\underline {Kinematics of 3C~129.}}
The galaxies associated with 3C~129 and 3C~129.1
(\cite{HillLongair1971} 1971) are approximately 0.7 and 1.05~mag
($B$-band) fainter than the brightest cluster member
(\cite{Sandage1975} 1975)
which is situated about $\sim$5$^{\prime}$ south of 3C~129.
Recently \cite{Nilssonetal2000} (2000, see also
\cite{ByrdValtonen1978} 1978) identified the brightest member
to be the cluster centre and they developed a detailed
kinematic model to explain the morphology of 3C~129,
after account has been taken of the buoyancy.
In the canonical picture,  a head-tail source is  FR I radio source
in a cluster of galaxies and the characteristic shape is explained
by the kinematics of the source and the properties of the beams
(\cite{Fabianetal2002} 2002).
Usually, the motion of the source is believed to be around the cluster center,
which is the dominant gravitational force in the cluster. The centre
of the cluster potential is also the location of X-ray emission due to the
trapping of the hot ICM and is also the location of
the brightest and most massive cD galaxy
which is often a powerful radio galaxy
(\cite{Riley1975} 1975, \cite{Bahcall1977} 1977 and \cite{Fabian1994} 1994).
The observations of 3C~129 and its surroundings suggest
that for this system, the picture is more complicated than the canonical
picture; in particular, the kinematical centre and the centre of the
X-ray emission for this cluster are different, which we discuss below:

\paragraph{\underline {Cluster centre:}}
The X-ray emission is elliptical ($\sim$10$^{\prime} \times \sim$6$^{\prime}$
with major in the east-west direction) and it peaks at and around the
location of two discrete sources: an extended source near the
core of the 3C~129.1 and a point source at 2$^{\prime}$ towards the
south-west of the 3C~129.1.
Although this point source seems to be at the centroid of X-ray emission,
the location of the cluster centre is assumed
to be between these two sources (\cite{Krawczynskietal2003} 2003).
If 3C~129.1 is at the centre of the cluster, one needs to explain why
it is not a cD galaxy and the dominant member of the cluster?
It is possible that the mass of its optical host could be
underestimated because of heavy obscuration
(\cite{Downes1980} 1980), hence
the centre of the cluster could not be determined from optical observations.
It is also possible that cluster is still in the process of growing via
cannibalism and becoming a cD galaxy
(\cite{McGlynnOstriker1980} 1980 and \cite{Dressler1984} 1984).
However, identification of 3C~129.1 with the cluster centre poses
problems for the kinematic model which explains the morphology of 3C~129.
Since, if 3C~129 is going around the cluster centre or falling into it,
it is not possible to generate the observed shape from kinematic models.

On the other hand, we probably do not understand the radio morphology
of 3C~129.1 and faint sources A, B and C may also be playing an important role.
The faint sources A and B are not associated with any known optical
sources.
We find that the source A has very steep spectrum
($\alpha_{240}^{610} \lesssim -$3.0$\pm$0.3) and could
be a radio relic source.
Whereas, the source C is coincident with the optical galaxy WEIN~047
and which is  also an IRAS source. This galaxy is detected in all
the bands ($J$, $H$, $K_{\rm s}$)
of 2~Metre All Sky Survey and is probably a starburst galaxy.
The radio spectral index, $\alpha_{240}^{1400}$~=~$-$1.51$\pm$0.11 is
typical of star-forming galaxies ({\it e.g.} M82:
\cite{Kellermannetal1971} 1971).
Careful look at the features in the maps of 3C~129.1 at 240 and
610~MHz reveal that its morphology is similar to most FR~I radio
galaxies, with bright regions on either side of the optical core
(see our Fig.~\ref{610mhzmap},
Fig.~5 (8.6~GHz) \& Fig.~6 (4.9~GHz) in \cite{Tayloretal2001} 2001, and
Fig.~2 (2.7~GHz) in \cite{Downes1980} 1980)
that fade with distance from the core.
Considering the fact that the source B is $\sim$3$^{\prime}$
south of the optical core and the farthermost north-eastern feature of
3C~129.1 is $\gtrsim$ 1$^{\prime}$ from it, and
their spectral indices, $\alpha_{240}^{610}$, are $\lesssim -$4.1$\pm$0.3
and $-$1.8$\pm$0.6 respectively, suggest that the spectrum steepens
along the jets and the steepening of spectrum is due to synchrotron
ageing of radio plasma.
Furthermore, the bright regions seen on either side of the optical
core and a sharp boundary on its south west side are either
because of episodic activity in the core or because
of buoyancy effects due to density inhomogeneities in the ICM.
Therefore, it seems that the source B is probably a part of
southern jet of 3C~129.1. This picture then suggests that
3C~129.1 is a wide-angle-tail source as suggested earlier by
\cite{Laneetal2002} (2002).

\section{Summary}
\label{summary}

We have presented the lowest frequency image of 3C~129
at 240~MHz. The important consequences of these observations
combined with our 610~MHz observations and VLA map of
\cite{Laneetal2002} (2002) at 330~MHz are as follows:

1. The morphology of 3C~129 is similar to that of twin-tailed
head-tail sources. Our both maps show curving tail detected
out to 22$^{\prime}$ from the head at 240~MHz and up to
$\sim$20$^{\prime}$ at 610~MHz.

2. 3C~129.1 by itself, is unlikely to be a twin-tailed source
and it being a head-tail source is also doubtful. Barring
projection effects, it could be similar to that of
bent-double sources. But if source~B is probably a part of
southern jet of 3C~129.1, then it could be a FR~I radio galaxy
which is a wide-angle-tail source as suggested earlier by
\cite{Laneetal2002} (2002).

3. The suggested scenario by
\cite{Laneetal2002} (2002) for the crosspiece is unlikely. Its
relatively flat spectrum ($\alpha$ = $-$1.04$\pm$ 0.22)
is not consistent with a fossil radio plasma revived
by 3C~129's bow shock.
We also do not believe \cite{JagersdeGrijp1983} (1983)
that crosspiece is one-sided ejection of plasmoids or beams.
Instead it seems to have formed due to irregularities and
inhomogeneities in the medium and/or perturbations in the jet.

4. The low frequency spectra between 240 and 610~MHz,
interpreted as a synchrotron ageing,
consist of four regimes along the tail:
(a) region between 0$^\prime$ and 2$^\prime$ where the
spectrum flattens from the head;
(b) region between 2$^\prime$ and 7$^\prime$ where the
spectrum steepens from $-$0.5 to $-$0.8;
(c) a region between 7$^\prime$ and 12$^\prime$ where
spectral index remains nearly constant; and
(d) a region beyond 12$^\prime$ where further steepening
occurs.
Such a behaviour was earlier noted by \cite{PacholczykScott1976} (1976).
We also confirm the result of \cite{PerleyEricson1979} (1979)
that straight spectra are seen at all points down the tail.

5. Assuming equipartition magnetic field and the
break frequency occurs at $\sim$240~MHz,
our age estimate of $\sim$200~Myr seems plausible
for 3C~129. This is in agreement to the age estimated by
\cite{Mileyetal1972} (1972).

6. There is some ambiguity as to the location of the cluster centre.
While it is more probable that it is near 3C~129.1, near the centroid
of the X-ray emission, locating it near the brightest galaxy in the
cluster has the advantage of explaining the kinematics of 3C~129.
Deep optical and infrared images would be useful in independently
establishing the cluster centre.

\begin{acknowledgements}
We thank the staff of the GMRT who have made these observations
possible. GMRT is run by the National Centre for Radio Astrophysics
of the Tata Institute of Fundamental Research.
We thank Wendy~Lane and her collaborators for the 330~MHz VLA map.
DVL thanks R.~Nityananda, S. Rawlings, S Roy and D.J. Saikia for
discussions and several useful comments.
This research has made use of the NASA/IPAC Extragalactic Database
which is operated by the Jet Propulsion Laboratory,
Caltech, under contract with the NASA.
\end{acknowledgements}


\begin{thebibliography}{1000}

\bibitem[Ananthakrishnan \& Rao] {AnanthRao2002}
         Ananthakrishnan, S., \& Rao, A.P., 2002, Proc. of
         a conf., Multi Colour Universe, held at the TIFR, India, p.~233
\bibitem[Baldwin \& Scott] {BaldwinScott1973}
         Baldwin, J.E., \& Scott, P.F. 1973, MNRAS, 165, 259
\bibitem[Baars et al.] {Baarsetal1977}
         Baars, J.W.M., Genzel, R., Pauliny-Toth, I.I.K., \& Witzel, A.
         1977, A\&A, 61, 99
\bibitem[Bahcall] {Bahcall1977}
         Bahcall, N.A. 1977, ARA\&A, 15, 505
\bibitem[Bennett] {Bennett1962}
         Bennett, A.S. 1962, Mem.R.A.S., 68, 163
\bibitem[Bologna et~al.] {Bolognaetal1969}
         Bologna J.M., McClain, E.F., \& Sloanaker, R.M. 1969, ApJ, 156, 815
\bibitem[van Breugel] {vanBreugel1982}
         van Breugel, M. 1982, A\&A, 110, 225
\bibitem[van Breugel \& Willis] {vanBreugelandWillis1981}
         van Breugel, M, \& Willis, A.G. 1981, A\&A, 96, 332
\bibitem[Byrd \& Valtonen] {ByrdValtonen1978}
         Byrd, G., \& Valtonen, M.J. 1978, ApJ, 221, 481 
\bibitem[Condon et~al.] {Condonetal1998}
         Condon, J.J., Cotton, W.D., Greisen, E.W., Yin, Q.F., Perley, R.A.,
         Taylor, G.B., \& Broderick, J.J. 1998, AJ, 115, 1693
\bibitem[Downes] {Downes1980}
         Downes, A. 1980, MNRAS, 190, 261
\bibitem[Downes] {Downes1984}
         Downes, A. 1984, MNRAS, 211, 215
\bibitem[Dressler] {Dressler1984}
         Dressler, A. 1984, ARA\&A, 22, 185
\bibitem[Fabian] {Fabian1994}
         Fabian, A.C. 1994, ARA\&A, 32, 277
\bibitem[Fabian et~al.] {Fabianetal2002}
         Fabian, A.C., Celotti, A., Blundell, K.M., Kassim, N.E.,
         \& Perley, R.A. 2002, MNRAS, 331, 369
\bibitem[Fanaroff \& Riley] {FanaroffRiley}
         Fanaroff, B.L., \& Riley J.M. 1974, MNRAS, 167P, 31
\bibitem[Gower et~al.] {Goweretal1967}
         Gower, J.F.R., Scott, P.F., \& Wills, D. 1967, Mem.RAS, 71, 49
\bibitem[Hill \& Longair] {HillLongair1971}
         Hill, J.M., \& Longair, M.S. 1971, MNRAS, 154, 125
\bibitem[Jaffe \& Perola] {JaffePerola1973}
         Jaffe, W.J., \& Perola, G.C. 1973, A\&A, 26, 423
\bibitem[J\"agers] {Jagers1987}
         J\"agers, W.J. 1987, A\&AS, 71, 603
\bibitem[J\"agers \& de~Grijp] {JagersdeGrijp1983}
         J\"agers, A.J., \& de~Grijp, M.H.K. 1983, A\&A 127, 235
\bibitem[Kardashev] {Kardashev1962} 
         Kardashev, N.S. 1962, Soviet Astron.--AJ, 6, 317
\bibitem[Kellermann et al.] {Kellermannetal1971} 
         Kellermann, K.I., \& Pauliny-Toth, I.I.K. 1971,
         Astrphys. Lett., 8, 153.
\bibitem[Komissarov \& Gubanov] {KomissarovGubanov1994}
         Komissarov, S.S., \& Gubanov, A.G. 1994, A\&A, 285, 27
\bibitem[Krawczynski et~al.] {Krawczynskietal2003}
         Krawczynski, H. Harris, D.E., Grossman, R., Lane, W.,
         Kassim, N., \& Willis, A.G. 2003, MNRAS, 345, 1255
\bibitem[van der Laan \& Perola] {vanderLaanPerola1969}
         van der Laan, H., \& Perola, G.C. 1969, A\&A, 3, 468
\bibitem[Lane et~al.] {Laneetal2002}
         Lane, W.M., Kassim, N.E., Ensslin, T.A., Harris, D.E., \& Perley, R.A.
         2002, AJ, 123, 2985
\bibitem[Macdonald et~al.] {Macdonaldetal1966} 
         Macdonald, G.H., Neville, A.C., \& Ryle, M. 1966, Nature, 211, 1241
\bibitem[McGlynn \& Ostriker] {McGlynnOstriker1980}
         McGlynn, T.A., \& Ostriker, J.P. 1980, ApJ, 241, 915 
\bibitem[Miley et al.] {Mileyetal1972} 
         Miley, G.K., Perola, G.C., van~der Kruit, P.C., \& van~der Laan, H.
         1972, Nature, 237, 269
\bibitem[Miley] {Miley1980}
         Miley, G.K. 1980, ARA\&A, 18, 165 
\bibitem[Nilsson et~al.] {Nilssonetal2000}
         Nilsson, K., Valtonen, M. Zheng, J.-Q., Byrd, G., Korhonen, H. \&
         Anderson, M.I. 2000, IAU Colloq. 174, ed. Valtonen, M.J. \& Flynn, C.
         2000, ASP Conf. Ser., (San Francsico: ASP), 209, 408
\bibitem[Owen et~al.] {Owenetal1979}
         Owen, F.N., Burns, J.O., Rudnick, L., \& Greisen, E.W.
         1979, ApJ, 229, L5
\bibitem[Pacholczyk] {Pacholczyk1970}
         Pacholczyk, A.G. 1970, Radio Astrophysics
         (San Francisco: W.E. Freeman \& Co.)
\bibitem[Pacholczyk \& Scott] {PacholczykScott1976}
         Pacholczyk, A.G., \& Scott, J.S. 1976, ApJ, 203, 313
\bibitem[Perley \& Erickson] {PerleyEricson1979}
         Perley, R.A., \& Erickson, W.C. 1979, ApJS, 41, 131
\bibitem[Riley] {Riley1973}
         Riley, J.M. 1973, MNRAS, 161, 167
\bibitem[Riley] {Riley1975}
         Riley, J.M. 1975, MNRAS, 170, 53
\bibitem[Rudnick \& Burns] {RudnickBurns1981}
         Rudnick, L., \& Burns, J.O. 1981, ApJ, 246, L69
\bibitem[Rudnick] {Rudnick2002}
         Rudnick, L. 2002, PASP, 114, 427
\bibitem[Ryle] {Ryle1960}
         Ryle, M. 1960, J. Instn elect. Engrs, 6, 14 
\bibitem[Sandage] {Sandage1975}
         Sandage, A.R. 1975, PASP, 87, 853
\bibitem[Slee et~al.] {Sleeetal2001}
         Slee, O.B., Roy, A.L., Murgia, M., Andernach, H., \& Ehle, M. 2001,
         AJ, 122, 1172
\bibitem[Slee et~al.] {Sleeetal1994}
         Slee, O.B., Roy, A.L., \& Savage, A. 1994, AuJPh, 47, 145 
\bibitem[Swarup et al.] {Swarupetal1991}
         Swarup, G., Ananthakrishnan, S., Kapahi, V.K., Rao, A.P.,
         Subrahmanya, C.R., \& Kulkarni, V.K. 1991, Cu. Sc., 60, 95
\bibitem[Taylor et~al.] {Tayloretal2001}
         Taylor, G.B., Govoni, F., Allen, S.A., \& Fabian, A.C. 2001, 
         MNRAS, 326, 2
\bibitem[White \& Becker] {WhiteBecker1992}
         White, R.L., \& Becker, R.H. 1992, ApJS, 79, 331
\bibitem[Wright et~al.] {Wrightetal1999}
         Wright, M., Dickel, J., Koralesky, B., \& Rudnick, L. 1999,
         ApJ, 518, 284

\end{thebibliography}
\end{document}